*Article*

# The crystal structure of $Pb_{10}(PO_4)_6O$ revisited: the evidence of superstructure

Sergey V. Krivovichev [1,2,*]

1 Nanomaterials Research Centre, Kola Science Centre, Russian Academy of Sciences, Fersmana 14, 184209 Apatity, Russia
2 Department of Crystallography, Institute of Earth Sciences, St. Petersburg State University, University Emb. 7/9, 199034 St. Petersburg, Russia
* Correspondence: s.krivovichev@ksc.ru

**Abstract:** The crystal structure of $Pb_{10}(PO_4)_6O$, the proposed matrix for the potential room-temperature superconductor LK-99, $Pb_{10-x}Cu_x(PO_4)_6O$ (x = 0.9-1.0), has been reinvestigated by single-crystal X-ray diffraction using crystals prepared by Merker and Wondratschek (*Z. Anorg. Allg. Chem.* **1960**, *306*, 25–29). The crystal structure is trigonal, $P\bar{3}$, $a$ = 9.8109(6), $c$ = 14.8403(12) Å, $V$ = 1237.06(15), $R_1$ = 0.0413 using 3456 unique observed reflections. The crystal structure of $Pb_{10}(PO_4)_6O$ is a superstructure with regard to the 'standard' $P6_3/m$ apatite structure type. The doubling of the $c$ parameter is induced by the ordering of 'additional' O′ atoms within the structure channels running parallel to the $c$ axis and centered at (00$z$). The O′ atoms form short bonds to the Pb1 atoms, resulting in splitting the Pb1 site into two, Pb1A and Pb1B. The structural distortions are further transmitted to the Pb phosphate framework formed by four Pb2 sites and PO4 groups. The structure data previously reported by Krivovichev and Burns (*Z. Kristallogr.* **2003**, *218*, 357-365) may either correspond to the $Pb_{10}(PO_4)_6O_x(OH)_{2-x}$ (x ~ 0.4) member of the $Pb_{10}(PO_4)_6O$ - $Pb_{10}(PO_4)_6(OH)_2$ solid solution series, or to the high-temperature polymorph of $Pb_{10}(PO_4)_6O$ (with the phase with doubled $c$ parameter being the low-temperature polymorph).

## 1. Introduction

In the series of recent preprints, S. Lee and co-authors [1-3] reported on the discovery of room-temperature superconductor LK-99 with the chemical formula $Pb_{10-x}Cu_x(PO_4)_6O$ (x = 0.9-1.0). The authors claimed that the superconducting properties of the material ($T_c$ = 400 K) are proved by the measurements of its resistivity, current, magnetic field, and the Meissner effect. Lee et al. point out that 'the superconductivity of LK-99 originates from minute structural distortion by a slight volume shrinkage (0.48%) … caused by $Cu^{2+}$ substitution of $Pb^{2+}$(2) ions in the insulating network of Pb(2)-phosphate' [1]. The stress caused by the substitution '…transfers to Pb(1) of the cylindrical column resulting in distortion of the cylindrical column interface, which creates superconducting quantum wells (SQWs) in the interface' [1]. The claims by Lee et al. [1-3] are currently under hot debates and detailed scrutiny by the international scientific community [4-8]. The density functional theory (DFT) studies on LK-99 have been performed by a number of groups [9-13], showing the interesting electronic features of LK-99, which itself raises important issues about its unique physical and structural properties. The structure models constructed for the DFT studies are based upon the available crystallographic data for $Pb_{10}(PO_4)_6O$ reported by Krivovichev and Burns [14] or the data on $Pb_{10}(PO_4)_6(OH)_2$ [15,16]. Both sets of data correspond to the high-symmetry apatite structure type with the $P6_3/m$ space group and the cell of $a$ ~ 9.85 and $c$ ~ 7.43 Å [17]. The proper understanding of the crystal structure of the matrix material $Pb_{10}(PO_4)_6O$ is of outmost importance for the elucidation of the mechanisms that generate superconductivity of LK-99.

The aim of the present paper is to re-consider the crystal structure of $Pb_{10}(PO_4)_6O$ in order to provide insight into its atomic nature and the character of structural distortions induced by its crystal-chemical modification. First, we provide some background information on $Pb_{10}(PO_4)_6O$, including historical notes and personal experience. Second, we present a new structure model of $Pb_{10}(PO_4)_6O$, based upon the experimental crystal-structure analysis. The crystal chemical analysis of the new model reveals its basic differences from the models assumed in the DFT modelling. Finally, we provide discussion of the possible mechanisms of the Pb-Cu substitution in LK-99 based upon the available crystal chemical information.

## 2. Background Information

'Oxypyromorphite' $Pb_{10}(PO_4)_6O$ has been known at least for more than 70 years. Rooksby [18] obtained it through crystallization of glass with respective chemical composition and provided the unit cell parameters $a$ = 9.75 and $c$ = 7.23 Å. The first detailed study of the synthesis and X-ray diffraction data of $Pb_{10}(PO_4)_6O$ were due to Merker and Wondratschek [19], who prepared the compound by crystallization from melt formed by mixing PbO and $NH_4H_2PO_4$ as initial reagents. The melting temperature of $Pb_{10}(PO_4)_6O$ was estimated as 967 °C (see Figure 3 in [19]) and the compound was reported as being melting incongruently [20]. Merker and Wondratschek [19] reported for the 'oxypyromorphite' a hexagonal unit cell with $a$ = 9.84 and $c$ = 14.86 Å, i.e., the superstructure of apatite with the doubled $c$ parameter. They pointed out that the differences between these parameters and the parameters reported by Rooksby [18] can be due to the possible mixed character of the sample studied by the latter ('Da es sehr viele Bleiverbindungen mit Apatit-Struktur und ahnlichen Gitterkonstanten gibt, die zudem weitgehend Mischkristalle bilden können, halten wir es nicht fur ausgeschlossen, das irgendeine andere dieser Verbindungen vorlag' [19]). Merker and Wondratschek [19] assigned the doubling of the $c$ parameter of their $Pb_{10}(PO_4)_6O$ to the ordering of 'additional' (not bonded to P) O atoms in the structure channels. Ito [21] suggested that the true formula of 'oxypyromorphite' can be written as $Pb^{2+}{}_9Pb^{4+}(PO_4)_6O_2$, in order to be in accord with the general formula of apatite-type compounds given as $A_5(TO_4)_3X$. However, this hypothesis was declined by Merker et al. [20], who demonstrated the validity of the $Pb^{2+}{}_{10}(PO_4)_6O$ composition.

In their study of the crystal structures lead oxide phosphates, Krivovichev and Burns [14] prepared and, for the first time, structurally characterized 'oxypyromorphite'. The crystals of $Pb_{10}(PO_4)_6O$ have been obtained by heating the mixture of 0.446 g PbO and 0.058 g $NH_4H_2PO_4$ to 950 °C in platinum crucibles, followed by cooling to 50 °C over 150 hours. The majority of the resulting reaction products consisted of elongated transparent crystals of $Pb_4O(PO_4)$ with a small quantity of isometric hexagonal crystals of $Pb_{10}(PO_4)_6O$. The single-crystal X-ray diffraction structure analysis [14] demonstrated that the latter compound crystallizes in the $P6_3/m$ space group with $a$ = 9.8650(3) and $c$ = 7.4306 Å. The structure model obtained is hereinafter denoted as $Pb_{10}(PO_4)_6O$-KB (after Krivovichev and Burns [14]).

After publication of the paper [14] in 2003, the author of this article had received an email message from Prof. Hans Wondratschek (1925-2014), who pointed out to the observation of Merker and Wondratschek [19] of the doubling of the $c$ parameter of $Pb_{10}(PO_4)_6O$ relative to the 'standard' apatite structure. Prof. Wondratschek also indicated the possibility that the phase studied by Krivovichev and Burns [14] could be, in fact, $Pb_5(PO_4)_3(OH)$, the conclusion that cannot be verified now due to the unavailability of the sample. After the suggestion of the re-study of the crystals of $Pb_{10}(PO_4)_6O$ using modern X-ray diffraction equipment, Prof. Wondratschek mailed to us the crystals prepared in 1968 by Prof. Günther Engel (University of Aalen, Germany). In addition, in 2006, Prof. Engel sent us additional samples for the further studies, which appeared to be structurally identical to those sent by Prof. Wondratschek. In this paper, the crystal structure model for $Pb_{10}(PO_4)_6O$ prepared by Wondratschek and Engel (hereinafter denoted as $Pb_{10}(PO_4)_6O$-WE).

## 3. Materials and Methods

The crystals of $Pb_{10}(PO_4)_6O$-WE used in this study have been obtained from Prof. Wondratschek and Prof. Engel as described above.

The single-crystal structure X-ray diffraction data have first been obtained by means of a STOE IPDS II X-ray diffractometer equipped with Image Plate area detector and operated at 50 kV and 40 mA. More than a hemisphere of three-dimensional data was collected using monochromatic MoK$_\alpha$ X-radiation, with frame widths of 2° in $\omega$, and with a 5 min count for each frame. The analysis of the reciprocal space revealed the presence of the reflections that correspond to the doubling of the $c$ unit-cell parameter, well in agreement with the observations by Merker and Wondratschek [19] (Fig. 1). The unit-cell parameters ($a$ = 9.832(1) and $c$ = 14.913(3) Å) were refined using least-squares techniques. The intensity data were integrated and corrected for Lorentz, polarization, and background effects using the STOE X-Red program. However, the attempts to solve and refine the crystal structure generally failed with the $R_1$ factor being always higher than 0.10.

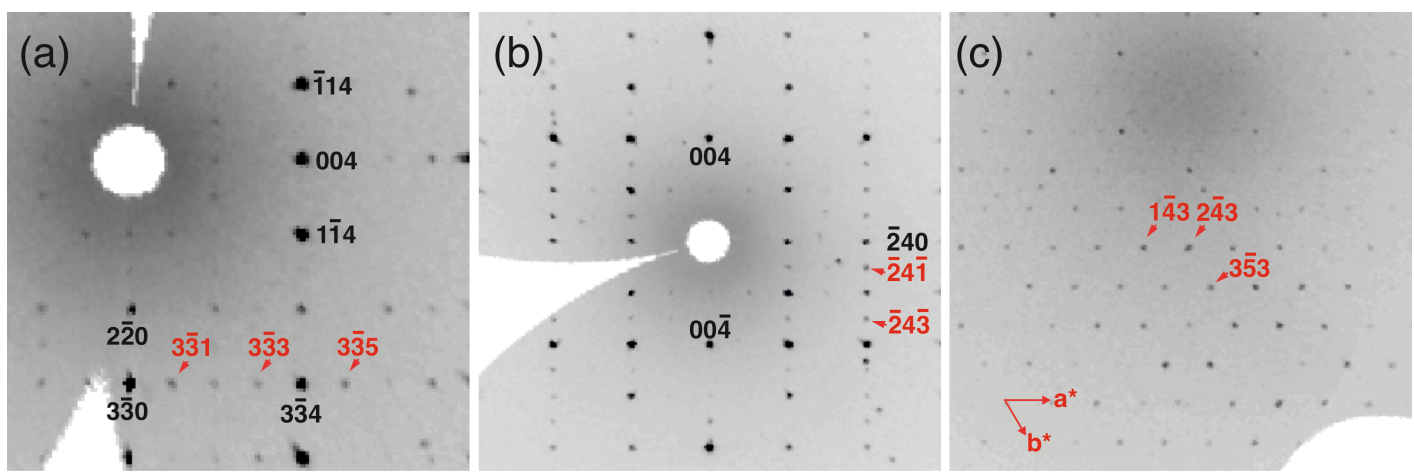

**Fig. 1**. Reconstructed sections of reciprocal diffraction space for $Pb_{10}(PO_4)_6O$-WE obtained using STOE IPDS II X-ray diffractometer: (a) section perpendicular to $[\bar{1}\bar{1}0]$; (b) section perpendicular to [100]; (c) the ($hk$3) section. The reflections doubling the $c$ parameter with respect to the 'standard' apatite structure are indicated by red arrows and red indices.

In order to obtain better data, the crystal of $Pb_{10}(PO_4)_6O$-WE was transferred to the Bruker Smart Apex II diffractometer at the Department of Crystallography, St. Petersburg State University, Russia. More than a hemisphere of X-ray diffraction data with the frame widths of 0.5° in $\omega$, and with 60 s spent counting for each frame were collected at room temperature using MoK$\alpha$ radiation. The indexing of the obtained reflections indicated the doubling of the $c$ parameter, in agreement with the observations by Merker and Wondratschek [19] and the Stoe IPDS II data. The data were integrated and corrected for absorption using the Bruker programs *APEX* and *XPREP*. The observed systematic absences were inconsistent with the presence of the $6_3$ screw axis characteristic for the 'standard' $P6_3/m$ space group typical for many apatite-type compounds. The subsequent attempts were undertaken to solve the crystal structure in the subgroups of $P6_3/m$, from which the only $P\bar{3}$ model provide more or less reasonable structure model with $R_1$ > 0.12. The twinning model was introduced using the $[100/010/00\bar{1}]$ matrix, which allowed the $R_1$ index to drop to ~0.05. The refinement resulted in the 0.503:0.497 ratio of the two twin components related to each other through the (001) mirror plane. The crystal structure was refined to $R_1$ = 0.0413 using 3456 unique observed reflections (Table 1). The *SHELX* program package was used for all structural calculations [22]. The final model included all atomic positional parameters, refinable weighting scheme of the structure factors and anisotropic-displacement parameters (ADPs) for all atoms. Anisotropic refinement of few O atoms resulted in physically unrealistic displacement parameters. In order to obtain physically reasonable model, the ADPs of the O atoms that are equivalent relative to the **c**/2 pseudo-translation were restrained to be equal. The final atomic coordinates, site occupation factors, bond-valence sums, and isotropic displacement parameters are given in Table 2, selected interatomic distances are listed in Table 3. The bond-valence sums for Pb and O atoms have been calculated using Pb-O bond-valence parameters from [23] and [24]; the P-O bond-valence parameters from [24] have been used to calculate the P BVSs. The CIF file and the list of observed and calculated structure factors are given as a Supplementary Information to this paper.

**Table 1.** Crystal data and structure refinement parameters for the $Pb_{10}(PO_4)_6O$-WE

| | |
|---|---|
| Temperature/K | 293(2) |
| Crystal system | trigonal |
| Space group | $P\bar{3}$ |
| $a$/Å | 9.8109(6) |
| $c$/Å | 14.8403(12) |
| Volume/Å$^3$ | 1237.06(15) |
| Z | 2 |
| $D_{calc}$, g/cm$^3$ | 7.135 |
| μ/mm$^{-1}$ | 68.270 |
| $F(000)$ | 2220 |
| Crystal size/mm$^3$ | 0.06 × 0.05 × 0.01 |
| Radiation | Mo$K\alpha$ (λ = 0.71073) |
| 2Θ range for data collection/° | 5.50 to 69.06 |
| Index ranges | -6 ≤ h ≤ 15, -15 ≤ k ≤ 11, -23 ≤ l ≤ 22 |
| Reflections collected | 11777 |
| Independent reflections | 3456 [$R_{int}$ = 0.0885, $R_{sigma}$ = 0.0955] |
| Data/restraints/parameters | 3456/0/105 |
| Goodness-of-fit on F$^2$ | 0.866 |
| Final $R$ indices [$I \geq 2\sigma(I)$] | $R_1$ = 0.0413, w$R_2$ = 0.0749 |
| Final $R$ indices [all data] | $R_1$ = 0.0544, w$R_2$ = 0.0771 |
| Largest diff. peak/hole / $e$ Å$^{-3}$ | 4.010/-4.537 |

**Table 2.** Atomic coordinates, bond-valence sums (BVS, v.u. = valence units) and isotropic displacement parameters (10$^{-4}$ Å$^2$) for $Pb_{10}(PO_4)_6O$-WE

| Site | BVS* | BVS** | x/a | y/b | z/c | $U_{iso}$ |
|---|---|---|---|---|---|---|
| Pb1A | 1.90 | 1.98 | 0.25369(5) | -0.00015(5) | -0.12920(4) | 0.01239(10) |
| Pb1B | 1.93 | 1.97 | -0.00445(6) | 0.22423(5) | 0.37576(6) | 0.02123(12) |
| Pb2A | 2.05 | 2.06 | 1/3 | 2/3 | 0.00402(8) | 0.0123(3) |
| Pb2B | 2.07 | 2.07 | 1/3 | 2/3 | 0.25395(7) | 0.0132(3) |
| Pb2C | 2.04 | 2.06 | 1/3 | 2/3 | 0.49982(8) | 0.0187(3) |
| Pb2D | 2.05 | 2.07 | 1/3 | 2/3 | -0.25587(7) | 0.0125(3) |
| P1A | 4.94 | 4.94 | 0.3740(3) | 0.4021(3) | 0.3698(3) | 0.0078(5) |
| P1B | 4.90 | 4.90 | 0.0230(3) | -0.3765(3) | -0.1215(3) | 0.0084(5) |
| O1A | 1.99 | 2.00 | -0.1578(9) | -0.4786(9) | -0.1223(9) | 0.0131(11) |
| O1B | 2.00 | 2.02 | 0.4953(10) | 0.3468(9) | 0.3692(9) | 0.0131(11) |
| O2A | 1.90 | 1.90 | 0.0807(14) | -0.2676(14) | -0.2034(6) | 0.0121(19) |

| | | | | | |
|---|---|---|---|---|---|
| O2B | 1.99 | 1.99 | 0.2777(14) | 0.3560(13) | 0.2826(6) | 0.0121(19) |
| O3A | 1.94 | 1.93 | 0.2533(14) | 0.3211(14) | 0.4471(6) | 0.021(2) |
| O3B | 1.85 | 1.85 | 0.0869(16) | -0.2716(17) | -0.0366(6) | 0.021(2) |
| O4A | 2.11 | 2.15 | 0.0949(9) | -0.4866(10) | -0.1233(9) | 0.0166(12) |
| O4B | 2.06 | 2.07 | 0.4563(10) | 0.5820(10) | 0.3809(9) | 0.0166(12) |
| O5A*** | 1.43 | 1.55 | 0 | 0 | 0.330(3) | 0.031(7) |
| O5B**** | 1.70 | 1.87 | 0 | 0 | 0.395(3) | 0.031(7) |

\* calculated using bond-valence parameters for Pb-O from [23].
\* calculated using bond-valence parameters for Pb-O from [24].
\*\*\* s.o.f. = 0.55(4).
\*\*\*\* s.o.f. = 0.45(4).

**Table 3.** Selected interatomic distances (Å) for the crystal structure of $Pb_{10}(PO_4)_6O$-WE

| | | | | |
|---|---|---|---|---|
| Pb1A-O4A | 2.234(8) | Pb2A-O1A | 2.505(11) | 3x |
| Pb1A-O2B | 2.451(10) | Pb2A-O4A | 2.791(11) | 3x |
| Pb1A-O2A | 2.554(12) | Pb2A-O3B | 2.835(13) | 3x |
| Pb1A-O3B | 2.638(10) | <Pb2A-O> | 2.710 | |
| Pb1A-O3B | 2.703(13) | | | |
| Pb1A-O1A | 2.801(8) | Pb2B-O4B | 2.588(11) | 3x |
| <Pb1A-O> | 2.564 | Pb2B-O1A | 2.647(11) | 3x |
| | | Pb2B-O2B | 2.847(11) | 3x |
| Pb1B-O5B | 2.240(6) | <Pb2B-O> | 2.694 | |
| Pb1B-O5A | 2.325(14) | | | |
| Pb1B-O3A | 2.453(12) | Pb2C-O4B | 2.503(11) | 3x |
| Pb1B-O4B | 2.731(8) | Pb2C-O1B | 2.616(11) | 3x |
| Pb1B-O3A | 2.749(10) | Pb2C-O3A | 3.172(12) | 3x |
| Pb1B-O2A | 2.757(10) | <Pb2C-O> | 2.764 | |
| Pb1B-O2B | 2.769(12) | | | |
| Pb1B-O1B | 3.057(8) | Pb2D-O1B | 2.428(11) | 3x |
| Pb1B-O3A | 3.175(12) | Pb2D-O4A | 2.843(11) | 3x |
| <Pb1B-O> | 2.695 | Pb2D-O2A | 2.961(11) | 3x |
| | | <Pb2D-O> | 2.744 | |
| | | | | |
| P1A-O2B | 1.532(11) | P1B-O2A | 1.527(11) | |
| P1A-O1B | 1.536(9) | P1B-O1A | 1.541(8) | |
| P1A-O4B | 1.539(9) | P1B-O3B | 1.548(12) | |
| P1A-O3A | 1.552(12) | P1B-O4A | 1.558(9) | |
| <P1A-O> | 1.540 | <P1B-O> | 1.544 | |

## 3. Results

Since $Pb_{10}(PO_4)_6O$-WE is superstructure with regard to the 'standard' $P6_3/m$ apatite structure type, it is reasonable to describe the structure peculiarities through the comparison of $Pb_{10}(PO_4)_6O$-WE ($P\bar{3}$, double $c$ parameter) with $Pb_{10}(PO_4)_6O$-KB ($P6_3/m$, single $c$ parameter). Both crystal structures are depicted in Fig. 2 in projections along the $c$ axes. In order to make the structure comparison easier, the atom sites in $Pb_{10}(PO_4)_6O$-WE have been numbered to be in maximal correspondence with the site numbering in $Pb_{10}(PO_4)_6O$-KB.

The transition between the KB and WE structure types can be described as consisting of two steps. First, the $P6_3/m$ space group changes to its maximal non-isomorphic space group $P\bar{3}$ with index 2. The transition is *translationengleiche*, that is, does not change the unit cell dimensions and is denoted as 't2' [25]. In the second step, the $c$ parameter is doubled [(**a**, **b**, **c**) → (**a**, **b**, 2**c**)] without space-group change, i.e., the transition is *klassengleiche* and is denoted as 'k2' (the space group $P\bar{3}$ (**a**, **b**, 2**c**) is a subgroup of index 2 of the space group $P\bar{3}$ (**a**, **b**, **c**)). Thus, the total transition from the KB and WE structure types can be described as a following sequence:

$P6_3/m$ (**a**, **b**, **c**) → t2 → $P\bar{3}$ (**a**, **b**, **c**) → k2 → $P\bar{3}$ (**a**, **b**, 2**c**),

where the intermediate $P\bar{3}$ (**a**, **b**, **c**) structure type is a hypothetical transitional structure. The relations between the Wyckoff sites occupied by Pb, P, and O atoms in the KB, transitional, and WE structure types are shown in Fig. 3. In order to distinguish between the O atoms of the PO$_4$ groups and the 'additional' O atoms not bonded to P, the latter are denoted as O'. Note that the occupancies of the O' sites in the $P6_3/m$ (**a**, **b**, **c**) and $P\bar{3}$ (**a**, **b**, **c**) structure types are equal to 0.25, whereas it is equal to 0.5 for the $P\bar{3}$ (**a**, **b**, 2**c**) structure type (Table 2).

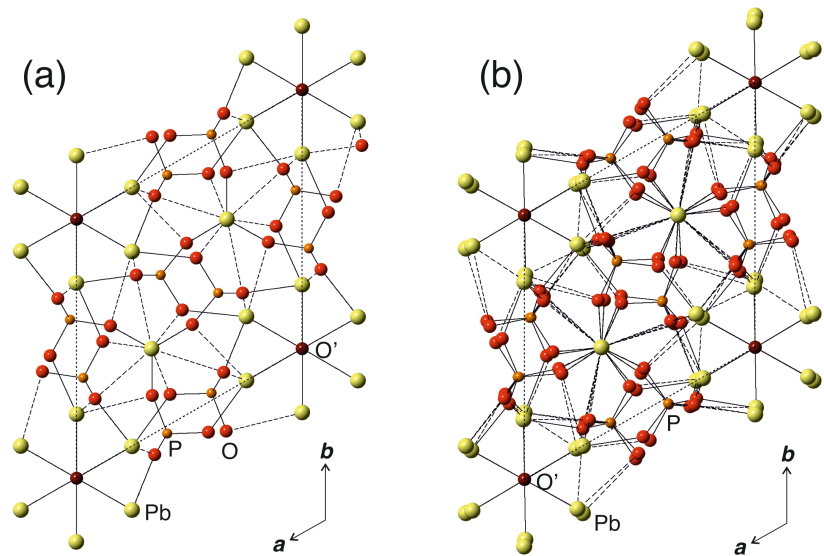

**Fig. 2**. The crystal structures of Pb$_{10}$(PO$_4$)$_6$O-KB (a) and Pb$_{10}$(PO$_4$)$_6$O-WE (b) in projections along the *c* axes. The Pb-O bonds shorter and longer than 2.5 Å are shown as solid and dashed lines, respectively.

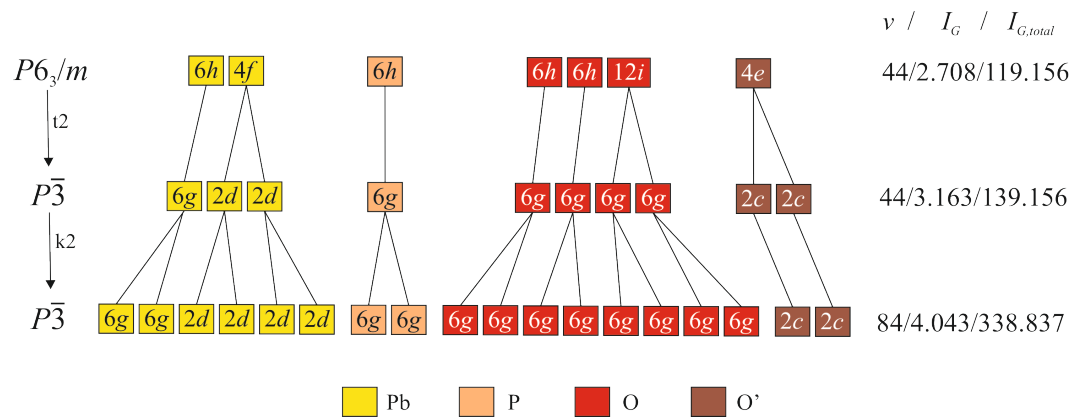

**Fig. 3**. Relations between the Wyckoff sites in the $P6_3/m$ (**a**, **b**, **c**) (Pb$_{10}$(PO$_4$)$_6$O-KB, $P\bar{3}$ (**a**, **b**, **c**) (hypothetical structure type) and $P\bar{3}$ (**a**, **b**, 2**c**) Pb$_{10}$(PO$_4$)$_6$O-WE. The complexity parameters for the three structure types are given on the right side of the Figure.

The diagram shown in Figure 3 indicates that, as a result of the imaginary Pb$_{10}$(PO$_4$)$_6$O-KB → Pb$_{10}$(PO$_4$)$_6$O-WE transition, the Pb1 site (6*h*) in the former is splitted into two 6*g* sites, Pb1A and Pb1B, in the latter. In the same manner, the Pb2 site (4*f*) is splitted into four 2*d* sites, Pb2A, Pb2B, Pb2C, and Pb2D. The coordination environments of Pb atoms within the coordination sphere of 3.2 Å as well as coordination of the O' atoms in the crystal structures of KB and WE are shown in Figure 4. It can be seen that the coordination of the Pb2 sites in the two structures are topologically identical and correspond to tricapped trigonal prisms. In contrast, the coordination of the Pb1 sites changes drastically. In $Pb_{10}(PO_4)_6O$-KB, the Pb1 site centers the basis of the $Pb1O_6$ pentagonal pyramid complemented by two Pb-O' bonds with the occupancy of the O' sites of 0.25. In $Pb_{10}(PO_4)_6O$-WE, there are two Pb1 sites. The Pb1A site forms the $Pb1AO_6$ pentagonal pyramid with no additional Pb-O' bonds and thus displays a typical coordination of $Pb^{2+}$ cations with stereochemically active $6s^2$ lone electron pairs [26]. The coordination of the Pb1B site is again the $Pb1BO_6$ pentagonal pyramid, but complemented by two Pb-O' (Pb1B–O5A and Pb1B-O5B) bonds and one longer Pb1B-O3A bond of 3.175 Å. Taking into account that Pb1A and Pb1B sites are connected by the **c**/2 pseudo-translation, the doubling of the *c* parameter in $Pb_{10}(PO_4)_6O$-WE is obviously due the ordering of the O' atoms in the structure channels centered by the [001] direction at (00*z*), which is in agreement with the prediction by Merker and Wondratschek [19].

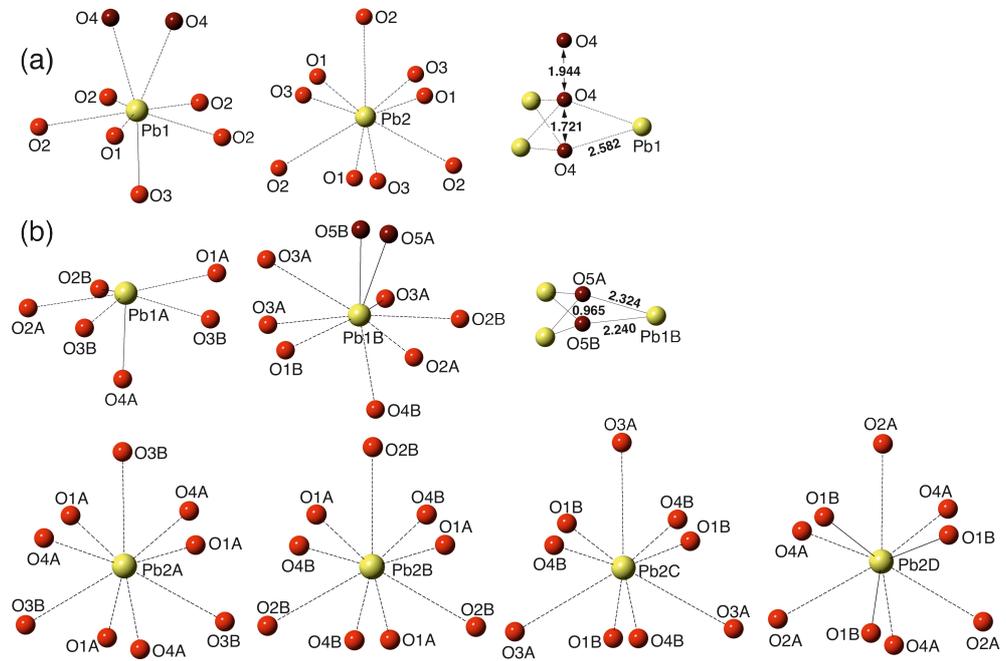

**Fig. 4**. Coordination polyhedra of Pb and O' sites in the crystal structures of $Pb_{10}(PO_4)_6O$-KB (a) and $Pb_{10}(PO_4)_6O$-WE (b). The Pb-O bonds shorter and longer than 2.5 Å are shown as solid and dashed lines, respectively.

The coordination of the O' sites in the two structures deserves special attention, especially in the light of the following discussion on the relations between the KB and WE phases (see section 4.2). In $Pb_{10}(PO_4)_6O$-KB, the O4 sites are arranged on the $6_3$ screw axis into columns with the O'-O' separations of 1.721 and 1.944 Å. Each O4 site is coordinated by three Pb1 atoms with the O4-Pb1 distance of 2.582 Å and the Pb1-O4-Pb1 angles of *ca*. 109.5°. The O4 atom is at the top of a trigonal pyramid with its basis formed by three Pb1 atoms. The bond-valence sum for the O4 atoms is 0.85 v.u., which is much lower than the expected value of 2.0 v.u. However, it should be taken into account that this coordination is averaged, since the site-occupation factor (s.o.f.) for the O4 site is 0.25.

In $Pb_{10}(PO_4)_6O$-WE, there are two O' sites, O5A and O5B, with s.o.f.s equal to 0.55 and 0.45, respectively. The O5A site forms three O5A-Pb1B bonds of 2.324 Å with the Pb1B-O5A-Pb1B angles of 111.8°. The O5B site is coordinated by three Pb1B atoms as well with the O5B-Pb1B bond lengths of 2.240 Å and the Pb1B-O5B-Pb1B angles of 118.4°, which means that the $O5B(Pb1B)_3$ configuration is very close to a planar triangle centered by the O5B site. Thus, the O′ sites in $Pb_{10}(PO_4)_6O$-WE are bonded much more tightly than those in $Pb_{10}(PO_4)_6O$-KB. The bond-valence sums for the O5A and O5B sites are 1.55 and 1.87 v.u., respectively. The O′-Pb bond lengths in $Pb_{10}(PO_4)_6O$-WE are in the ranges typical for the Pb oxysalt compounds with 'additional' O atoms [26].

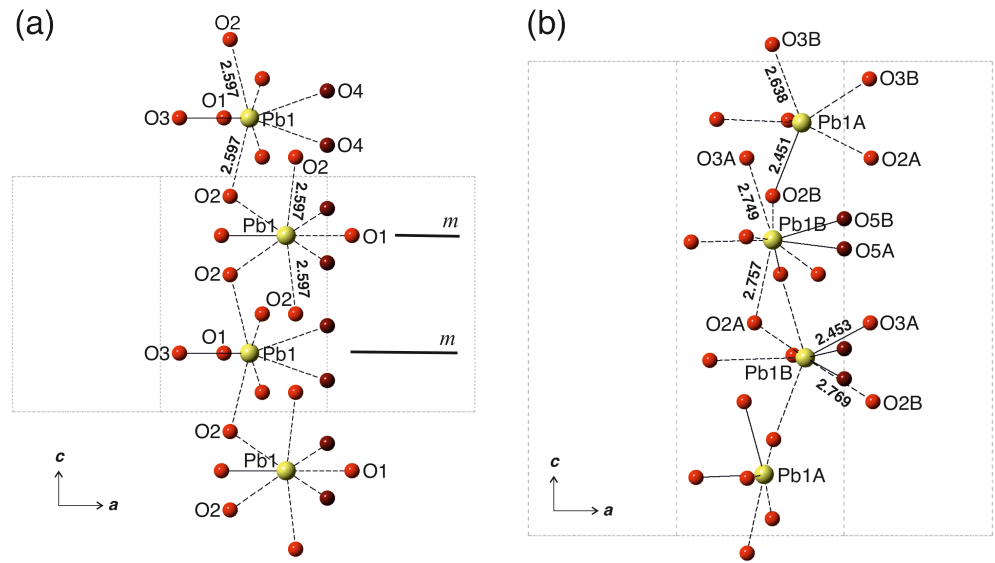

**Fig. 5**. The chain of the $Pb1O_n$ coordination polyhedra in $Pb_{10}(PO_4)_6O$-KB (a) and $Pb_{10}(PO_4)_6O$-WE (b). The Pb-O bonds shorter and longer than 2.5 Å are shown as solid and dashed lines, respectively.

## 4. Discussion

### 4.1. Symmetry considerations

The imaginary $Pb_{10}(PO_4)_6O$-KB → $Pb_{10}(PO_4)_6O$-WE transition is associated with the removal of the *m* mirror plane running in KB through the Pb1, O1 and O3 atoms at $z = ¼$. Fig. 5 shows projections of the columns of $Pb1O_n$ coordination polyhedra in both structures. In the crystal structure of $Pb_{10}(PO_4)_6O$-KB, the O′ (O4) atoms are arranged regularly along the *c* axis and the *m* plane is conserved. In $Pb_{10}(PO_4)_6O$-WE, the O′ atoms are in more ordered configuration, which results in the splitting of Pb1 site into two sites, Pb1A not bonded to O′, and Pb1B bonded to O′. The reasons for the disappearance of the *m* plane in WE are especially clearly seen from the comparison of the Pb1B-O3A and Pb1B-O2B bonds that are related by the mirror plane in the parent structure of KB. In the crystal structure of WE, the Pb1B-O3A and Pb1B-O2B bond lengths are 2.453 and 2.769 Å, respectively, with the difference larger than 0.3 Å.

The absence of the *m* plane in WE can also be illustrated by considering the column of tricapped $Pb2O_9$ prisms in both structures (Fig. 6). Again, the symmetry breaking can be illustrated by the comparison of the Pb2D-O1A and Pb2C-O1A bond lengths (2.428 and 2.617 Å, respectively), which differ by ~0.2 Å. The level of distortion of the substructure of the Pb2 polyhedra in the structure of WE is lower than that for the Pb1 substructure, which is reasonable, taking into account that Pb1A and Pb1B atoms are in the channels occupied by O′ atoms. Thus, the ordering of the O′ atoms in the structure channels results in the distortions of the coordination environments of the Pb1 atoms, which is transmitted to the Pb phosphate substructure formed by Pb2 atoms.

### 4.2. The relations between KB and WE

The relations between $Pb_{10}(PO_4)_6O$-KB and $Pb_{10}(PO_4)_6O$-WE is the crucial problem for the understanding of the crystal structure of $Pb_{10}(PO_4)_6O$ and its Cu-substituted derivatives. There are two possible solutions to this question: (i) $Pb_{10}(PO_4)_6O$-KB is in fact $Pb_{10}(PO_4)_6(OH)_2$; (ii) $Pb_{10}(PO_4)_6O$-KB and $Pb_{10}(PO_4)_6O$-WE are two polymorphs of Pb₁₀(PO₄)₆O separated by the order-disorder phase transition. We consider the two possibilities separately.

Table 4 provides unit-cell parameters reported in the literature for 'oxypyromorphite' $Pb_{10}(PO_4)_6O$, hydroxypyromorphite $Pb_{10}(PO_4)_6(OH)_2$, and their Cu-substituted varieties, along with some details of the coordination of O' atoms ($O^{2-}$ or $(OH)^-$ anions).

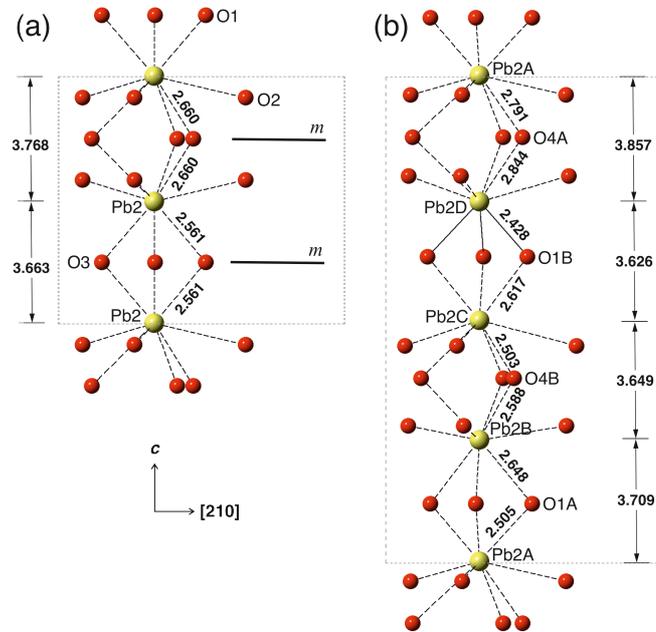

**Fig. 6**. The chains of the Pb₃O₉ tricapped trigonal prisms in $Pb_{10}(PO_4)_6O$-KB (a) and $Pb_{10}(PO_4)_6O$-WE (b). The Pb-O bonds shorter and longer than 2.5 Å are shown as solid and dashed lines, respectively.

**Table 4.** Unit-cell parameters and coordination of O' atoms in $Pb_{10}(PO_4)_6O$, $Pb_{10}(PO_4)_6(OH)_2$, and their Cu-substituted varieties

| Phase | $a$, Å | $c$, Å | $V/Z$, Å³ | O' coordination | O'-Pb, Å | Ref. |
|---|---|---|---|---|---|---|
| $Pb_{10}(PO_4)_6O$-WE | 9.84 | 14.86 | 623.0 | - | - | 19 |
| $Pb_{10}(PO_4)_6O$-WE | 9.811 | 14.840 | 618.5 | trigonal | 2.240/2.325 3x | this work |
| $Pb_{10}(PO_4)_6O$-KB | 9.865 | 7.431 | 626.3 | trigonal | 2.582 3x | 14 |
| $Pb_{10}(PO_4)_6(OH)_2$ | 9.866 | 7.426 | 626.0 | trigonal | 2.896 3x | 15 |
| $Pb_{10}(PO_4)_6(OH)_2$ | 9.883 | 7.441 | 629.4 | trigonal | 2.588 3x | 16 |
| $Pb_{10}(PO_4)_6(OH)_2$ | 9.774 | 7.291 | 603.2 | octahedral | 2.926 6x | 27 |
| $Pb_{10}(PO_4)_6(OH)_2$ | 9.871 | 7.427 | 626.7 | - | - | 28 |
| LK-99 | 9.843 | 7.428 | 623.2 | - | - | 3 |
| $Pb_8Cu_2(PO_4)_6(OH)_2$ | 9.870 | 7.398 | 624.1 | - | - | 28 |
| $Pb_6Cu_4(PO_4)_6(OH)_2$ | 9.868 | 7.392 | 623.4 | - | - | 28 |
| $Pb_4Cu_6(PO_4)_6(OH)_2$ | 9.866 | 7.383 | 622.3 | - | - | 28 |

First, we note that the $V/Z$ value for $Pb_{10}(PO_4)_6O$-WE (618.5 Å³) is smaller than that for LK-99 (623.2 Å³), which means that incorporation of $Cu^{2+}$ into $Pb_{10}(PO_4)_6O$ does not result in the volume shrinkage (S. Lee et al. [3] based their conclusion on the V/Z value for $Pb_{10}(PO_4)_6O$-KB, which is, indeed, higher than that for LK-99).

Second, it can be clearly seen that, in terms of the unit-cell parameters, $Pb_{10}(PO_4)_6O$-KB is most closely related to $Pb_{10}(PO_4)_6(OH)_2$ as reported by Brueckner et al. [15] and Kim et al. [16] (the results by Barinova et al. [27] are obviously for another Pb phosphate apatite, most propably $Pb_5(PO_4)_3F$). The O' coordination in $Pb_{10}(PO_4)_6O$-KB is almost identical to the coordination of OH group in $Pb_{10}(PO_4)_6(OH)_2$ as reported by Kim et al. [16] from neutron diffraction studies.

The free refinement of the X-ray diffraction data used by Krivovichev and Burns [14] indicated the occupancy of the O4 site equal to 0.40, whereas 0.25 and 0.50 are expected for $O^{2-}$ and $(OH)^-$ anions, respectively. Therefore, it is quite likely that the crystal studied in [14] was in fact an intermediate member of the $Pb_{10}(PO_4)_6O$ - $Pb_{10}(PO_4)_6(OH)_2$ solid solution, that is, the compound $Pb_{10}(PO_4)_6O_x(OH)_{2-x}$ with x ~ 0.4. However, taking into account that $Pb_{10}(PO_4)_6O$-KB was prepared by heating the mixture of PbO and $NH_4H_2PO_4$ to 950 °C, the scenario of its formation should involve incorporation of $H_2O$ either from air or from the traces of $NH_4H_2PO_4$.

Another possibility is that $Pb_{10}(PO_4)_6O$-KB and $Pb_{10}(PO_4)_6O$-WE are two polymorphs of $Pb_{10}(PO_4)_6O$ separated by order-disorder phase transition. To check this hypothesis, one should investigate behavior of the crystal structure of $Pb_{10}(PO_4)_6O$-WE (which is supposed to be a more ordered low-temperature modification) under high-temperature conditions.

*4.3. Structural complexity*

The information-based structural complexity parameters of the $Pb_{10}(PO_4)_6O$-KB and $Pb_{10}(PO_4)_6O$-WE structure types as well as of the intermediate hypothetic structure type with the space group $P\bar{3}$ (**a**, **b**, **c**) have been calculated using the following equations [29,30]:

$$I_G = -\sum_{i=1}^{k} p_i \log_2 p_i \text{ (bit/atom)}, \quad (1)$$

$$I_{G,total} = -v\sum_{i=1}^{k} p_i \log_2 p_i \text{ (bit/cell)}, \quad (2)$$

where $k$ is the number of different crystallographic orbits (Wyckoff sites) in the structure and $p_i$ is the random choice probability for an atom from the $i$th crystallographic orbit, that is:

$$p_i = m_i/v, \quad (3)$$

where $m_i$ is a multiplicity of a crystallographic orbit (i.e., the number of atoms of a specific Wyckoff site in the reduced unit cell) and $v$ is the total number of atoms in the reduced unit cell.

The $v/I_G/I_{G,total}$ parameters for the three structure types under consideration are given on the right side of Figure 3. The symmetry breaking in the sequence of transitions $P6_3/m$ (**a**, **b**, **c**) → t2 → $P\bar{3}$ (**a**, **b**, **c**) → k2 → $P\bar{3}$ (**a**, **b**, 2**c**) corresponds to the gradual increase of structural complexity, both in terms of information amounts per atom and per reduced unit cell. If $Pb_{10}(PO_4)_6O$-KB and $Pb_{10}(PO_4)_6O$-WE are polymorphs, the relation of their complexities corresponds to the general observation that high-temperature polymorphs are simpler than their low-temperature counterparts.

*4.4. On the mechanism of the Pb-Cu substitution in 'oxypyromorphite'*

The incorporation of $Cu^{2+}$ into the crystal structure of $Pb_{10}(PO_4)_6O$ ('oxypyromorphite') was claimed to be crucial for the superconductivity of LK-99. The current DFT calculations are based upon the assumption that $Cu^{2+}$ is accommodated into the Pb2 site of the $P6_3/m$ structure [9-13]. Such an accommodation should be accompanied by considerable local disorder as observed in some complex Pb-Cu oxides [31-33]. The disorder is due to the different electronic structure of $Pb^{2+}$ and $Cu^{2+}$ cations. The $Pb^{2+}$ cation possesses a $6s^2$ lone electron pair, which may be stereochemically active in the presence of strong Lewis bases [26]. In contrast, $Cu^{2+}$ cations in oxygen environments experience the Jahn-Teller distortion with characteristic planar-square coordination complemented by longer Cu-O bonds [34-36]. In some cases, the Cu-Pb substitution results in the splitting of the cation site into separate sites, as it was observed, e.g., for prewittite, $KPb_{1.5}ZnCu_6(SeO_3)_2O_2Cl_{10}$ [37], where the disordered $Cu^{2+}$ and $Pb^{2+}$ cations are separated by 1.04 Å, each having its specific coordination environment. The coordination of $Cu^{2+}$ ions in LK-99 should deserve special attention, especially in the case if the unusual superconducting properties of this material are confirmed.

Ben Moussa et al. [28] studied the incorporation of $Cu^{2+}$ ions into the crystal structure of hydroxypyromorphite and obtained the series of phases with the general composition $Pb_{10-x}Cu_x(PO_4)_6(OH)_2$ with x = 2, 4, 6. The unit-cell parameters for these phases are given in Table 4. The incorporation of Cu into $Pb_{10}(PO_4)_6(OH)_2$ results in the general shrinkage of the unit cell with the *c* parameter being the most sensitive to the value of x.

## 5. Conclusions

In conclusion, we reported the crystal structure of $Pb_{10}(PO_4)_6O$ with doubled *c* unit-cell parameter and the space group $P\bar{3}$. The symmetry breaking relative to the 'standard' apatite $P6_3/m$ space group is the result of ordering of 'additional' O' atoms in the structure channels running along the *c* axis and centered at (00z). Our results show that the crystal structure of LK-99 may appear to be more complex than it is assumed in the recent DFT calculations, which should be taken into account in further studies of this material.

**Supplementary Materials:** The following supporting information can be downloaded at: CIF and structure factors file for $Pb_{10}(PO_4)_6O$-WE.

**Funding:** This research was funded by the Russian Science Foundation, grant 19-17-00038.

**Data Availability Statement:** The structural data for $Pb_{10}(PO_4)_6O$-WE are available upon request..

**Acknowledgments:** This paper is dedicated to the memory of Professor Hans Wondratschek (1925-2014), an extraordinary crystallographer and great man, for his outstanding contributions to the crystal chemistry of apatite-type minerals and inorganic compounds.

**Conflicts of Interest:** The author declares no conflicts of interest.